EUROPEAN LABORATORY FOR PARTICLE PHYSICS

CERN - LHC DIVISION

# Measurements of the Secondary Electron Emission of Some Insulators

*J. BARNARD, I. BOJKO, N. HILLERET*

# Measurements of the secondary electron emission of some insulators.

John BARNARD, Iouri BOJKO and Noël HILLERET

*LHC-VAC, CERN, 1211 Geneva 23, Switzerland*

Charging up the surface of an insulator after beam impact can lead either to reverse sign of field between the surface and collector of electrons for case of thick sample or appearance of very high internal field for thin films. Both situations discard correct measurements of secondary electron emission (SEE) and can be avoided via reducing the beam dose. The single pulse method with pulse duration of order of tens microseconds has been used. The beam pulsing was carried out by means of an analog switch introduced in deflection plate circuit which toggles its output between "beam on" and "beam off" voltages depending on level of a digital pulse. The error in measuring the beam current for insulators with high value of SEE was significantly reduced due to the use for this purpose a titanium sample having low value of the SEE with DC method applied. Results obtained for some not coated insulators show considerable increase of the SEE after baking out at $350^0$C what could be explained by the change of work function. Titanium coatings on alumina exhibit results close to the ones for pure titanium and could be considered as an effective antimultipactor coating.

## Introduction

The multipactoring phenomenon remains until now one of the dominant mechanism limiting achievement of the high gradient fields in microwave cavities. Ceramic windows and gaps being an essential part of the construction of the cavities take a particular place in the multipactoring occurring in the cavities since the secondary electron emission coefficient (SEEC) of insulators is usually very high what causes high electron load to the ceramic. This high electron load leads to desorption of gases from the surface of the ceramic, pressure rise, appearance of discharge and finally window breakdown[1].
Other to the cavities the similar problem occurs in high voltage separators on the ceramic fixing deflection plates. Therefore it is an important question of the correct choice of ceramics having low value of SEEC for RF devices . The present paper aims to overview problems arising during secondary electron emission measurements on insulators, describe an experimental set-up and present results of the SEEC measurements for non-coated and coated insulators .

## Charging up the surface of an insulator

After an beam impact on the surface of a sample of insulator having an area $A$ and thickness $d$ (Fig, 1a) with following removal (for SEEC>1) or adsorption (SEEC<1) of electrons, a charged spot of area $S$ appears on the surface of the sample, where $S$ is approximately the beam cross section. If for an energy of the incident beam $E_p$ the insulator has a value of SEEC equal $\delta_p$ than

$$\delta_p = 1 - \frac{I_s}{I_b} \qquad (1)$$

where $I_B$ is the beam current and $I_S$ is current induced on the sample. For the duration of the beam impact with the insulator $\Delta\tau$ the charge of the spot will be

$$q_p = I_s \cdot \Delta\tau = (\delta_p - 1) \cdot I_b \cdot \Delta\tau \qquad (2)$$

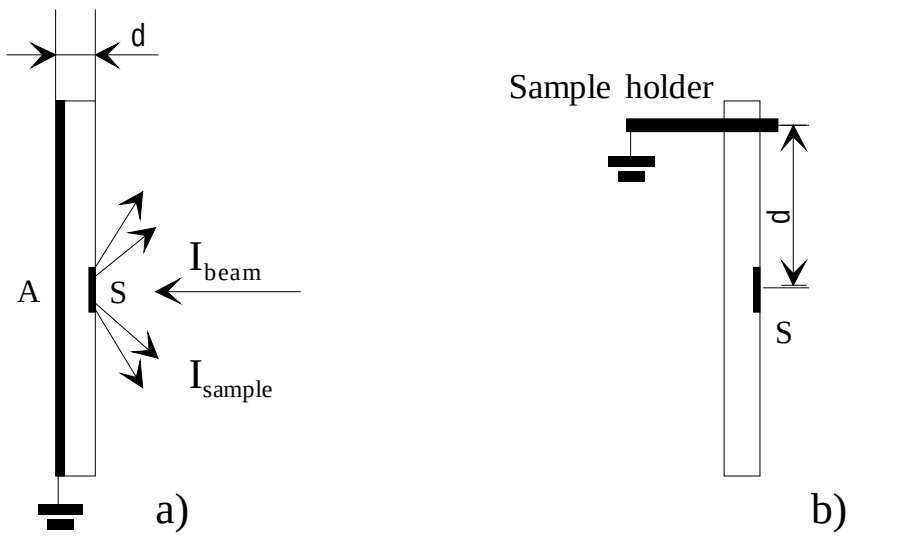

*Fig.1. Schematic diagram of the interaction of the electron beam with an insulator surface.*
*a) thickness d of the sample defines voltage difference*
*b) no thickness influence on voltage difference defined here by distance between beam spot and the sample holder.*

Assuming that the back side of the sample is metallized and has a ground potential one can think this assembly as a parallel-plate capacitor having spacing $d$, charge $q_p$ and plates of areas $A$ and $S$. If the charge distribution over the spot $S$ is suggested to be uniform, one can then use standard expressions derived for a parallel-plate capacitor having the both plates of area $S$. Thus, the voltage difference between plates $S$ and $A$ will be defined as

$$V_p = \frac{q_p \cdot d}{\kappa \cdot \varepsilon_0 \cdot S} \qquad (3)$$

where $\varepsilon_0$ is permittivity constant and $\kappa$ is a dielectric constant of the insulator material.
Combining the equations (2) and (3) one obtains

$$V_p = \frac{(\delta_p - 1) \cdot I_b \cdot \Delta\tau \cdot d}{\kappa \cdot \varepsilon_0 \cdot S} \qquad (4)^1$$

This appearing surface potential $V_p$ due to a beam impact puts serious problems on the measurements of the secondary electron emission for insulators. One can distinguish two different mechanisms disturbing the measurements and depending on the value of the SEEC and the thickness of the sample.

First mechanism relates to the case of a thick sample. On a typical energy distribution of the collected electrons from a sample due to a beam impact with energy $E_p$ one can see two principal peaks - elastic and "true" secondaries (Fig. 2). The elastic peak occurs at energy

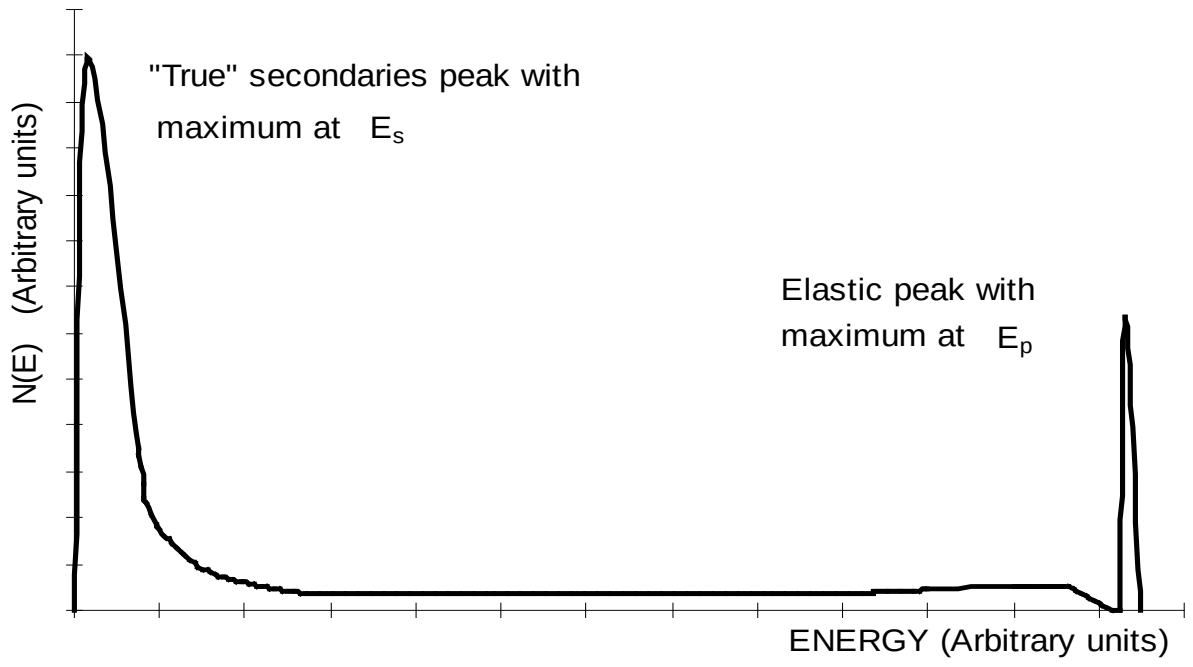

*Fig.2. Schematic energy distribution of collected electrons for the primary beam with energy $E_p$.*

$E_p$ and the true secondaries peak occurs at energy $E_s$. For insulators the value of $E_s$ is of order of few electron-volts. If the acquired surface potential $V_p$ becomes more positive than the potential $V_c$ of the collector of the electrons, the difference of the potentials $V_p$ - $V_c$ will act as a potential barrier for the secondary electrons - those having energies less than $V_p$ - $V_c$ can not escape from the surface and will be thus lost for the measurements. Obviously the more $V_p$ - $V_c$ the more secondaries lost and the more the error of measurements.

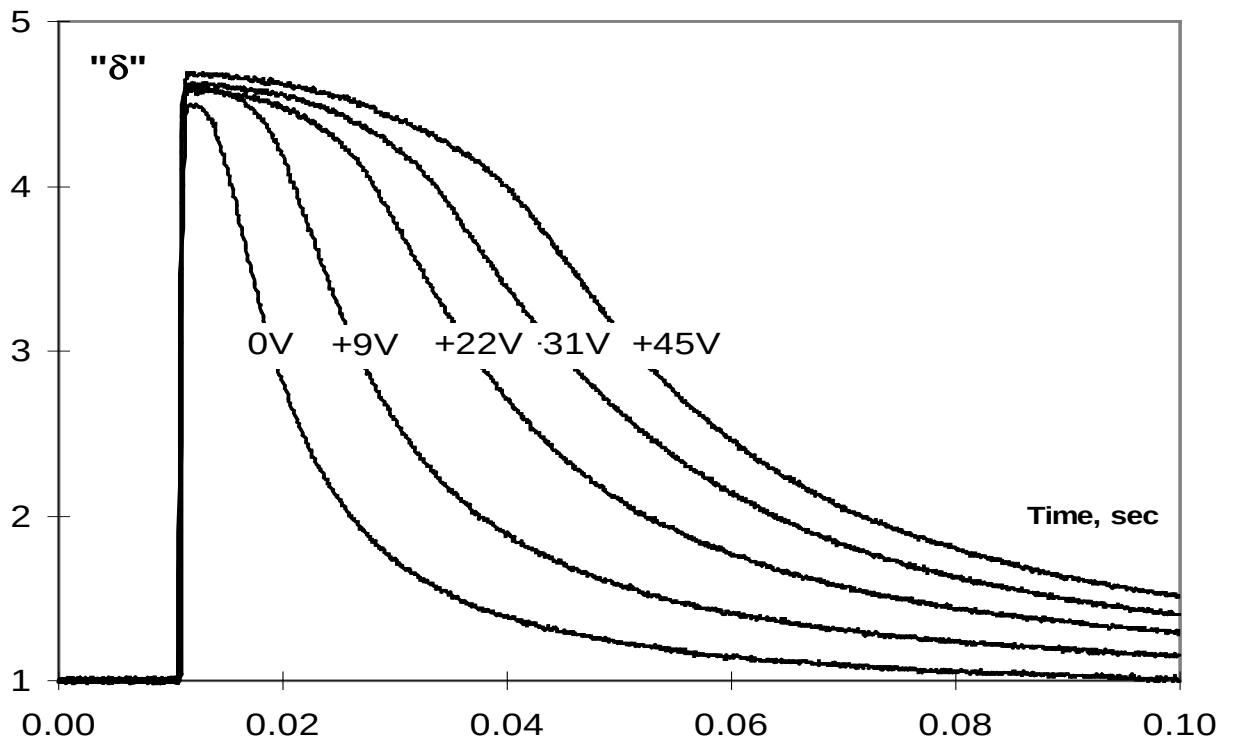

*Fig.3 Behavior of the SEEC for an alumina sample as function of the bombardment time for different values of the collector potential.*

This mechanism can be illustrated by an experiment results of which are shown in Fig.3 An alumina sample with thickness of 0.2mm was bombarded with 3keV energy beam having current of 1.1•10$^{-9}$A. Behavior of the SEEC in function of time has been measured for different values of the collector potential. One can clearly see on the graph that the more positive values of the collector potential the more bombarding time remains available for keeping an error in determining SEEC in tolerable level.

Second mechanism deals with the case of a very thin sample when during irradiation of an insulator with electron beam, a field gradient $V_p / d$ across insulator can reach a value causing the escape of additional secondaries . The field appearing across the insulator after beginning of the bombardment accelerates secondary electrons during their travel to the surface mainly via pores of a substrate to an energy sufficient to liberate new electrons. These electrons create new ones by ionization and a Townsend-type avalanche results [2]. A stable value of the electron yield could be reached when the surface potential approaches to the collector potential. But before this stable situation is reached, the field gradient $V_p / d$ across insulator can become so high that results appearance of field emission which sometimes remains even after the stopping of the bombardment. Both, Townsend-type avalanche and SEE caused by field emission are illustrated in Fig.4 where 700 monolayers of argon deposited on a copper target at 4.2K were bombarded with 3keV electron beam having current 2.9•10$^{-9}$A. The potential of the collector during this experiment was kept at +45V.

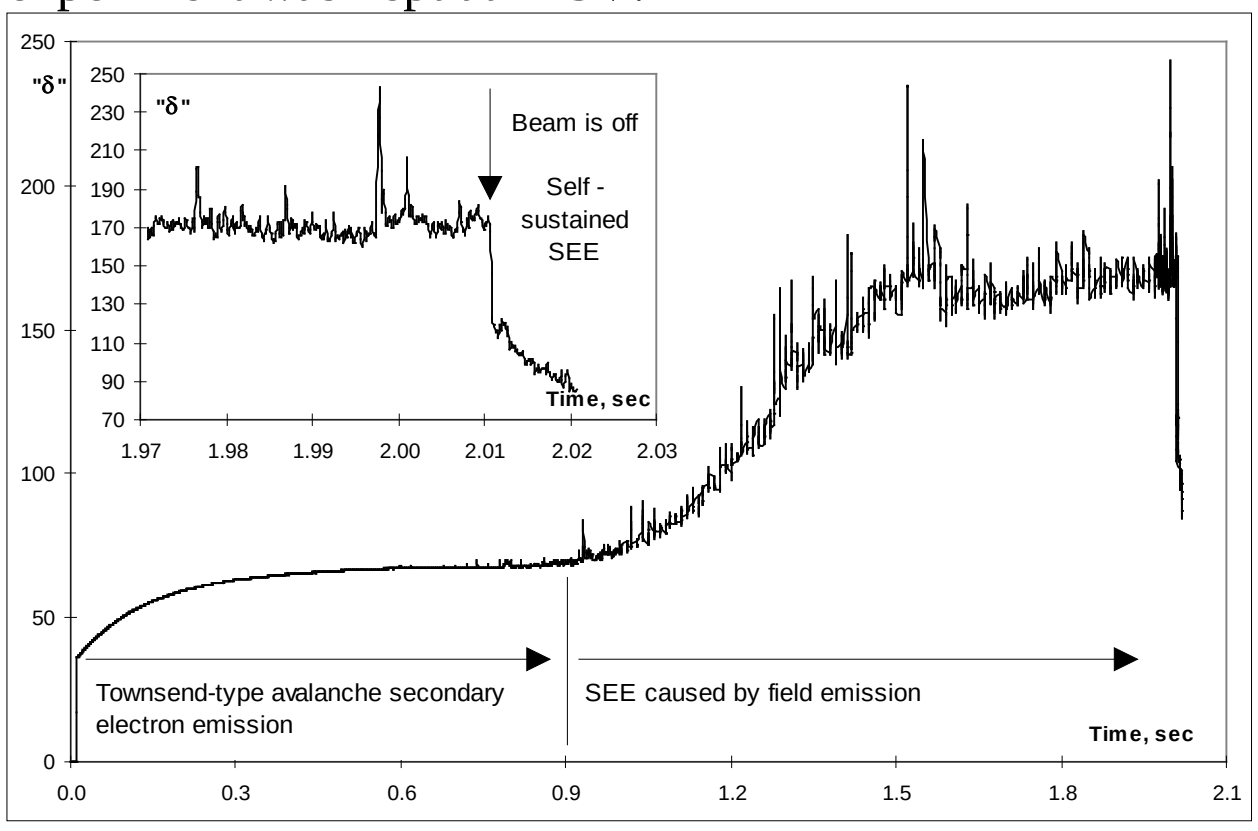

*Fig.4. Secondary electron yield as function of the bombardment time obtained on 700 monolayers of argon deposited on a copper target at 4.2K.*

These examples put in evidence an importance for keeping the surface potential $V_p$ as small as possible for correct measurements of the SEE of insulators. The equation (4) shows the ways for this - increasing the beam diameter and reducing the beam current, the sample thickness and the beam pulse duration. The substantial gain can be obtained by reducing the beam pulse duration - so-called "single pulse method".

Estimations of value of the electron dose necessary for the measurements of the SEEC of insulators with a reasonable error can be done on the basis of the measurements of previous authors. Thus for a crystal of KCl and for the incident beam energy $E_p$ =1500eV the SEEC was found as $\delta_p \approx 13$[3]. The admissible value of the electron dose before the surface potential reaches the one of the collector can be obtained from the equation (4) for a sample of thickness $d$ =1mm, the beam diameter $D_{beam}$ =3mm, the dielectric constant of KCl $\kappa$ =2.4[4] and the collector potential $V_c$ =+45V as

$$I_b \cdot \Delta\tau = \frac{V_c \cdot \kappa \cdot S \cdot \varepsilon_0}{(\delta_p - 1) \cdot d} \approx 5.5 \bullet 10^{-13}\,\mathrm{C}$$

---

[1]It is necessary to note that the thickness $d$ of the insulator appearing in the equations (2)-(4) means a distance between the surface beam spot and a nearest point with a fixed potential. Thus, for an example depicted on the Fig.2 b) where a sample holder is such a point, the sample radius should be taken as a "thickness" $d$ .

Assuming the beam current to be $I_b = 5.5 \bullet 10^{-9}$A one obtains $\Delta\tau$ =100μsec and $I_s = 6.6 \bullet 10^{-8}$A.

These values point out to a direction for SEEC measurements for insulators - keeping surface charge in a tolerable level via decreasing the primary beam pulse duration. This way was employed for the first time for energy distribution measurements[5] and was afterwards called "single-pulse method".

The task of measurements such values of currents with the beam pulse duration of order of a hundred of microseconds can be simplified by measuring the value of the charge accumulated by the sample. If a RC filter with the constant time much more than the beam pulse duration will be introduced in a input current circuit then the current signal in the circuit will correspond accumulation of charge what gives after integrating of the signal the value of the charge acquired by the sample during the beam pulse[3,6,7]. The result divided on the pulse duration gives the value of the current.

## Experimental set-up and procedure of measurements

The measurements of the SEEC for insulators were carried out on an apparatus depicted on Fig. 5.

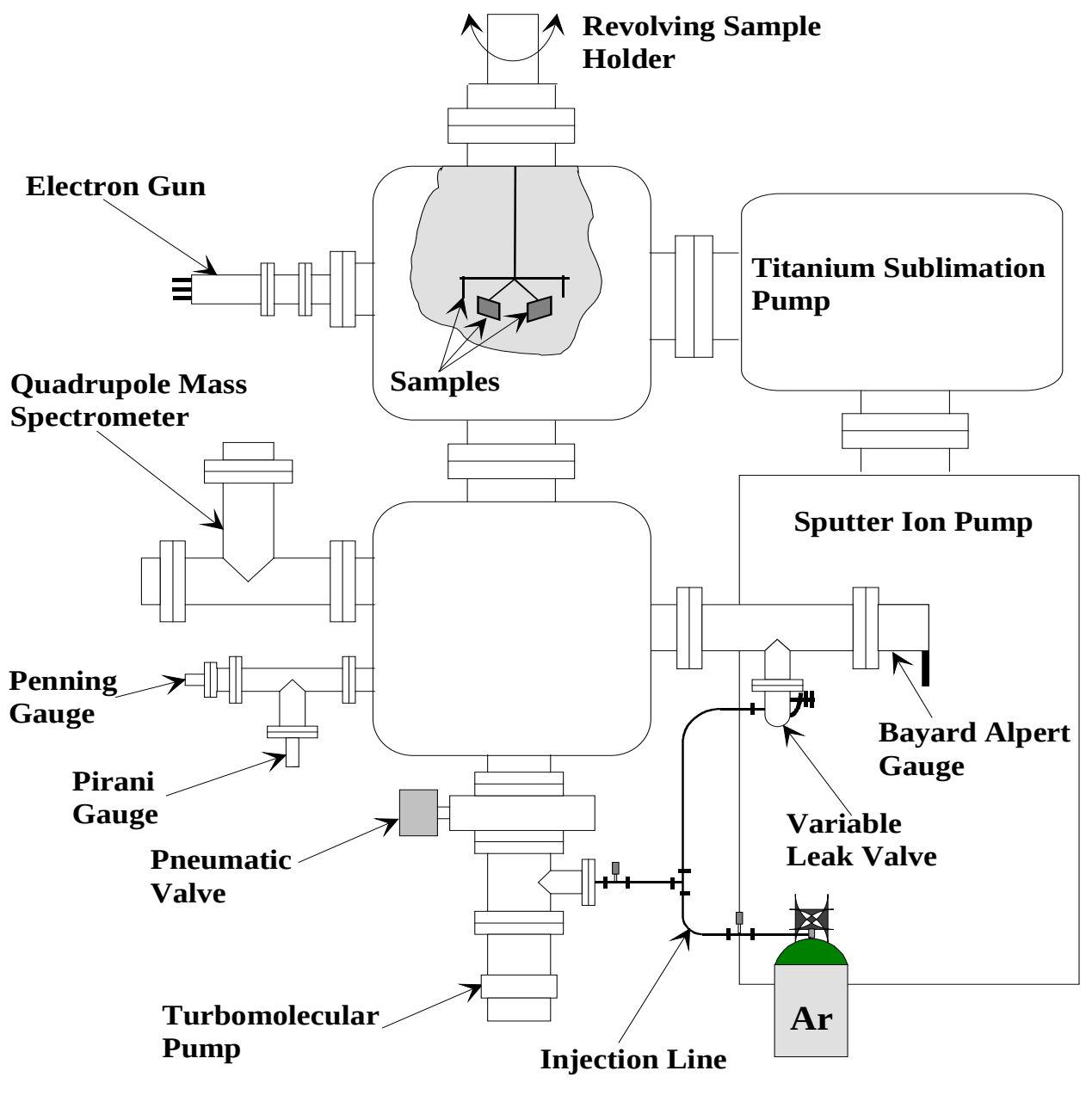

*Fig. 5. Schematic view of the vacuum system.*

The vacuum system consists basically of the vacuum vessel, the vacuum pumps and the gauges for total and partial pressure measurements. The vacuum vessel is made of stainless steel and can be baked out at 350$^0$C. Pumping is provided by a sputter ion pump and a titanium sublimation pump. A turbomolecular pump, which is backed by an oil-sealed rotary vane pump, serves mainly to evacuate the system from atmospheric pressure down to $10^{-7}$torr and it is valved off afterwards.

An experimental arrangement for SEY measurements is shown on Fig.6 and it consists of an electron gun able to accelerate electrons until 3keV, a collector of electrons (cage), a revolving sample holder, a filament for stabilization of the surface potential, beam driving and current measuring electronic equipment. The cage was biased at +45V relatively ground in order to prevent escape of secondary electrons from its surface. The set-up can be loaded with 14 samples at once.

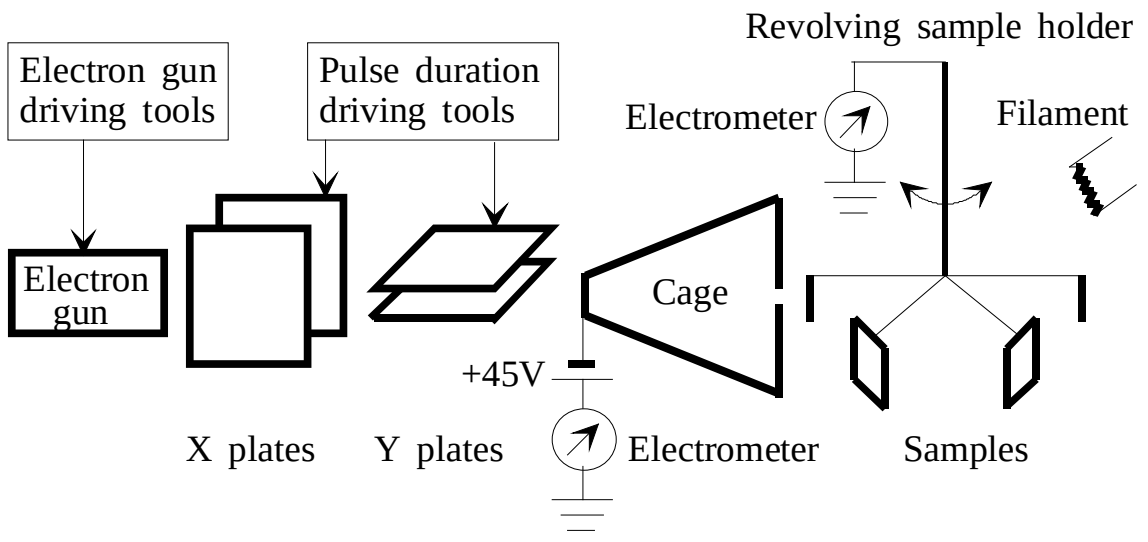

*Fig. 6. Diagram of the experimental arrangement.*

If $I_c$ is the current measured on the cage then the value of SEEC for a sample one can calculate as

$$\delta = \frac{I_c}{I_c + I_s} \qquad (6)$$

where

$$I_c + I_s \equiv I_b \qquad (7)$$

represents the beam current. The equation (6) can be rewritten in terms of electron doses acquired by the cage and the sample as

$$\delta = \frac{\int_0^\tau I_c(\tau)d\tau}{\int_0^\tau I_c(\tau)d\tau + \int_0^\tau I_s(\tau)d\tau} \qquad (8)$$

A problem arising from usage the equations (6)-(8) is such that the error of the measurements of the beam current can become unacceptably high when dealing with high values of SEEC. In fact, taking into account right signs the equation (7) for the SEEC>1 changes to

$$I_c - I_s \equiv I_b \qquad (7')$$

Since both the sample and cage currents are of order of $\delta \cdot I_b$ for the SEEC>>1 it leads to the fact that an error in determination of the beam current could reach values as high as $2 \cdot \delta \cdot \theta$ , where $\theta$ is an error of measurements for the sample and the cage currents. Thus, if $\theta$ is of order of 2% then the error in determination of the beam current can reach 40% for an insulator having SEEC=10. This problem can be simply overcome by doing the measurements of the beam current on a metal sample "dummy" which has much lower value of the SEEC.

Another advantage of this is that the DC method can be implemented what improves considerably signal-to-noise ratio during the measurements. Correspondingly the equation (8) is replaced by an equation

$$\delta = \frac{\int_0^{\tau} I_c(\tau)d\tau}{I_b \Delta\tau} \qquad (9)$$

where $\Delta\tau$ is the beam pulse duration.

Since after each beam impact the surface of an insulator becomes charged it is necessary to relax the charge before doing the next measurement. It was carried out by flooding the surface with low energy electrons from a filament whilst keeping the sample and the filament at ground potential.

A 12 bits fast scanning board MIO-16E-2 from National Instruments with 2 μsec analog-to-digit conversion time was used to measure the currents and to drive the beam. Beam pulsing was carried out by means of an analog switch introduced in a deflection plate circuit. The switch toggles its output between "Beam-off" and "Beam on" voltages depending on level of a digital pulse coming from a counter of the board (Fig. 7). As effect the duration of the digital pulse defines the duration of the beam pulse. At the same time the digital pulse also triggers acquisition for the current channels.

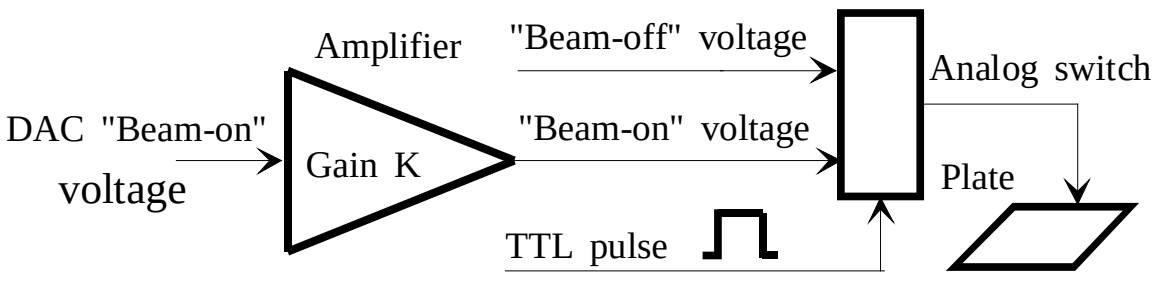

*Fig. 7. The beam pulse driving electronic scheme*

Current amplifiers Keithley 427 were used as electrometers. The filter time constant of the electrometers was chosen to be 0.3msec that corresponded to 1.4msec as the total current response time. Since the total analog-to-digit conversion time for two current channels makes up 4 μsec, about 350 points of the measurements for each of the channels $I_c$ and $I_s$ can be taken. This number is great enough to perform precise integration.

## Results

This chapter presents results of SEEC measurements for not coated and coated insulators. All samples have been measured in "as received" state and after baking out at $350^0$C during 24 hours. The pressures in the vacuum chamber were about $5.0 \cdot 10^{-8}$torr during the measurements in "as received" state and $2.0 \cdot 10^{-10}$torr after baking out.

The beam current during the measurements was chosen to be between $1.0 \cdot 10^{-9}$A and $2.0 \cdot 10^{-9}$A with the beam diameter being about 2mm. The beam current was measured with the help of a titanium sample having low value of SEEC.

Before putting in the chamber the back side of each sample was coated with gold. The beam dose for each sample during measurements varied depending of its value of SEEC. In fact a few preliminary attempts for every sample have been carried out in order to define the maximal allowable dose leading to the best signal-to-noise ratio. The lowest beam dose used during the SEEC measurements for insulators was about $2 \cdot 10^{-14}$C. It corresponded to the beam pulse duration of order of 20μsec.

<u>*Not coated insulators*</u>

In total 6 samples of different insulators have been investigated - quartz, zyranox, sapphire, alumina of 94% purity, alumina of 97.6% purity and pure alumina. All samples were of thickness of 0.2mm.

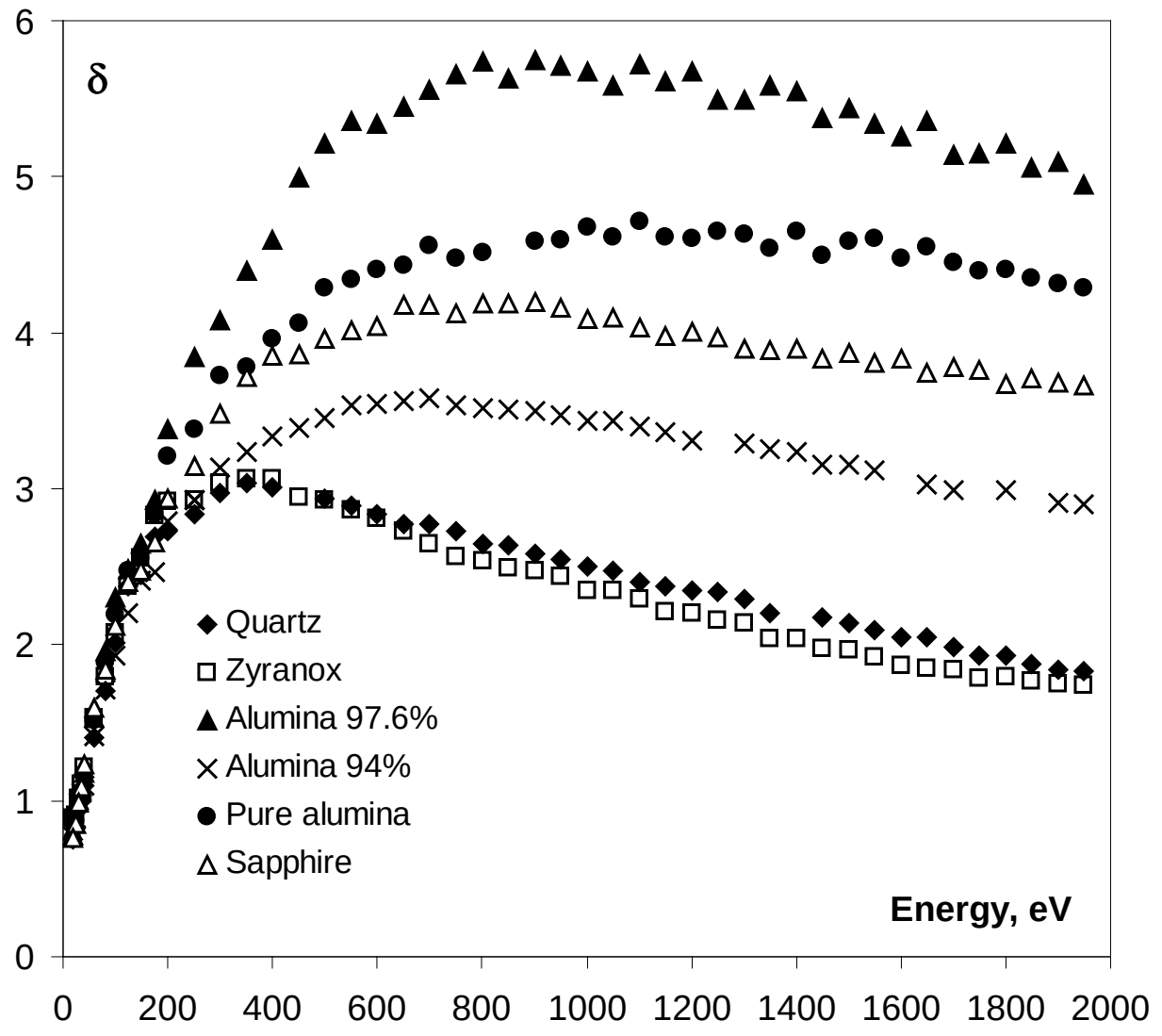

*Fig. 8. SEEC for insulators in "as received" state.*

The results of the SEEC measurements for these insulators in "as received" state and after baking out are presented correspondingly on Figures 8 and 9.

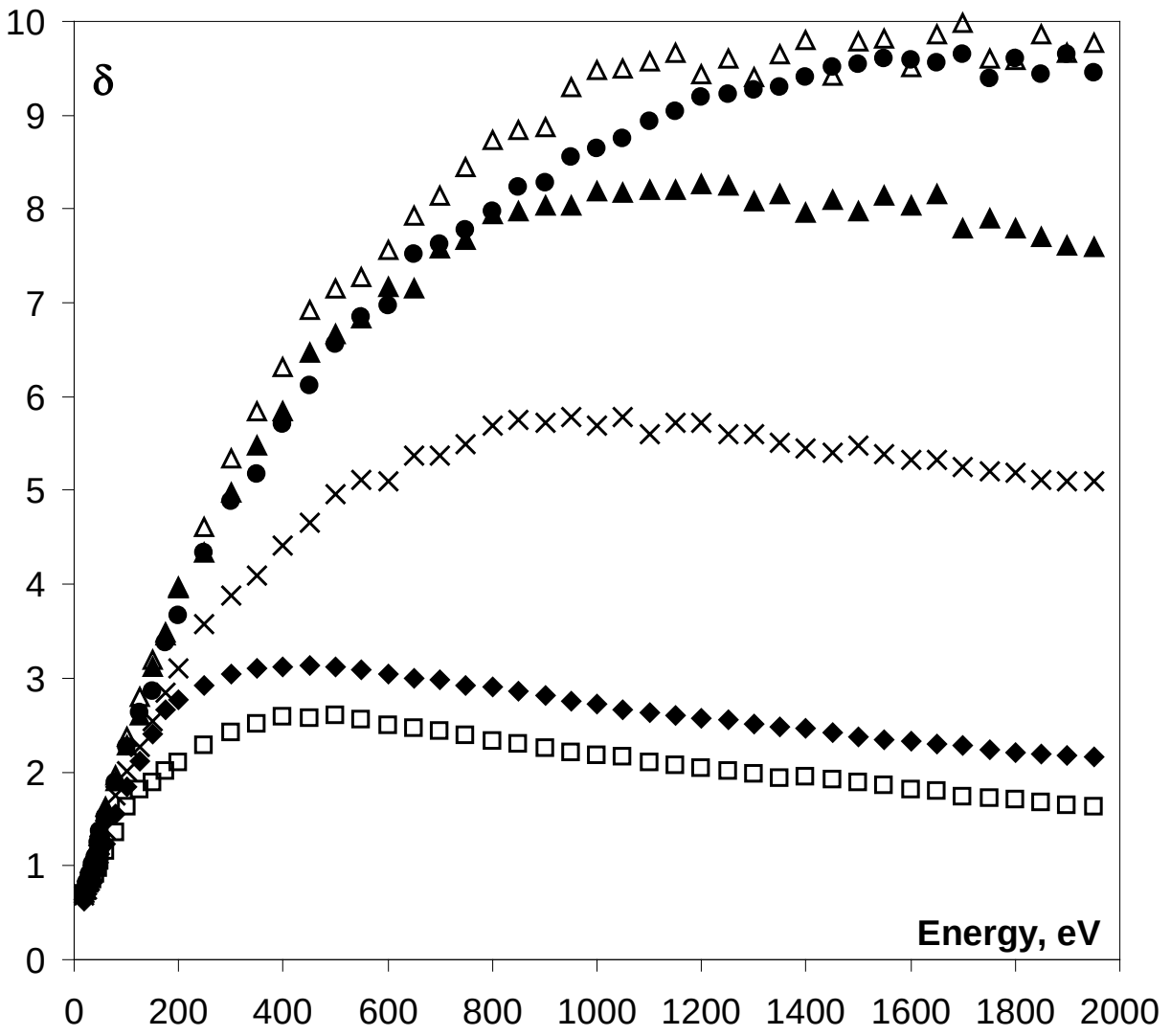

*Fig. 9. SEEC for insulators after baking out at $350^0$C*

The maximal values of SEEC for each sample in "as received" state $\delta_{m0}$ and after baking and the values of energy where this maximum is attained are summarized in Table 1.

**Table 1. Maximal values of the SEEC and their corresponding values of the primary energy.**

| Crystal | “as received” state | | after baking out 350$^0$C | |
|---|---|---|---|---|
| | $d_{max}$ | $E_{max}$ | $d_{max}$ | $E_{max}$ |
| Quartz | **3** | **370** | **3.15** | **405** |
| Zyranox | **3.06** | 335 | 2.6 | **470** |
| Alumina 97.6% | **5.7** | **935** | **8.2** | **1150** |
| Alumina 94% | **3.55** | **695** | **5.75** | **1000** |
| Pure alumina | **4.6** | **1090** | **...** | **...** |
| Sapphire | **4.2** | **775** | **...** | **...** |

*Coated insulators*

It is already a proven way to use coatings for preventing the multipactoring occurring in klystron and coupler windows[8,9,10]. Titanium based and air oxidized chrome coatings seem to be the most often used for such purposes.

*Titanium coated alumina samples.*

Two series of titanium coated alumina samples have been tested. First series has been prepared at DESY and it had 3 samples of dimensions 30mm•30mm and thickness of 0.6mm. The samples have been coated with pure titanium of different thickness which was controlled via measuring the surface resistance between two opposite corners of a sample. The surface resistance for these samples have made up

- sample 11/1 - 244MΩ
- sample 11 - 372MΩ
- sample 12 - 800MΩ

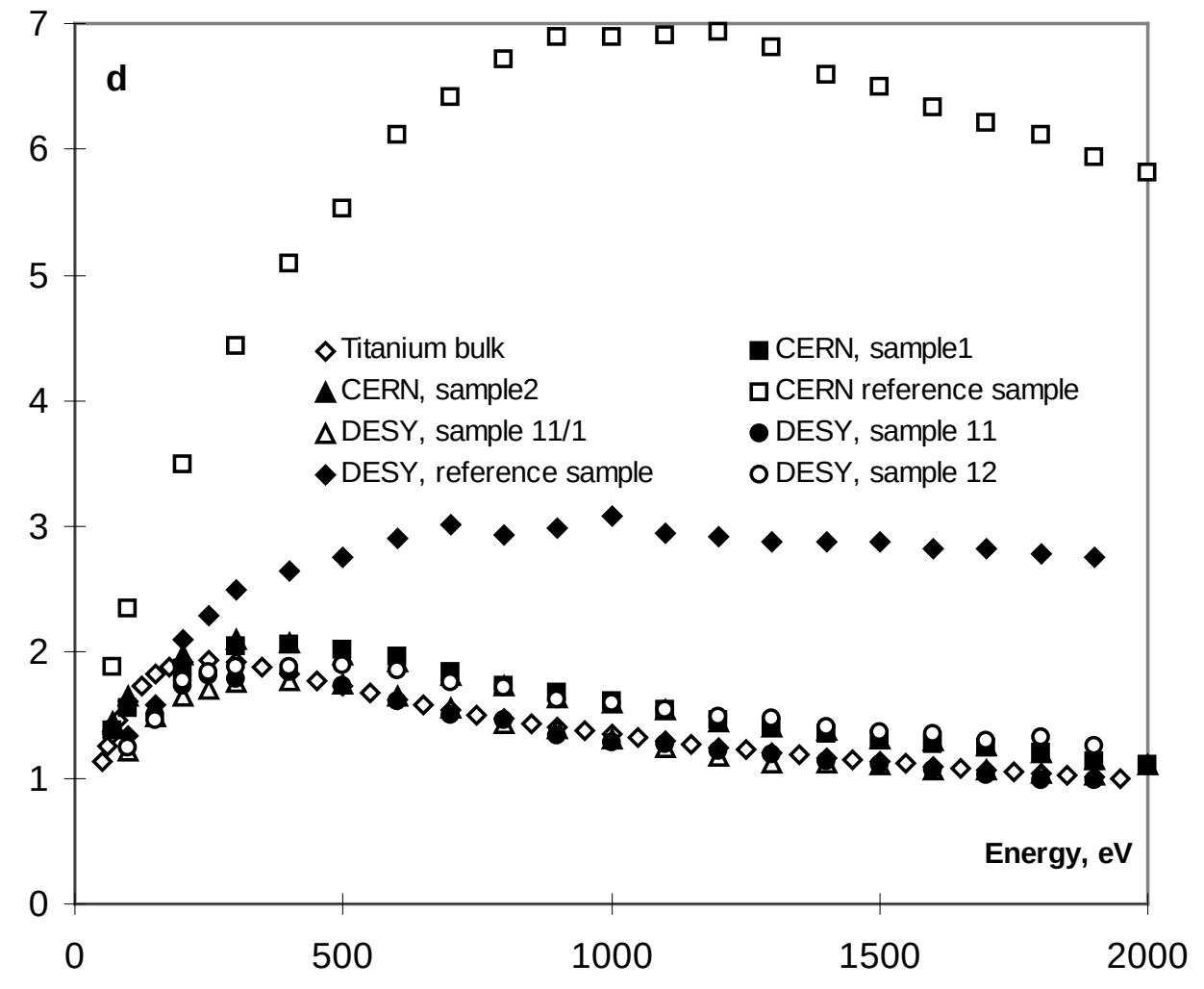

*Fig. 10. SEEC for titanium coated alumina samples in “as received” state.*

The second series of titanium coated alumina samples have been prepared at CERN. Two alumina samples of dimensions 30mm•10mm and thickness of 0.2mm have been put on a internal wall of an alumina coupler window. The material used for the samples was the same as for the window. The assembly was afterwards sputtered with titanium. The thickness of the titanium coating is estimated as 100-150Å .

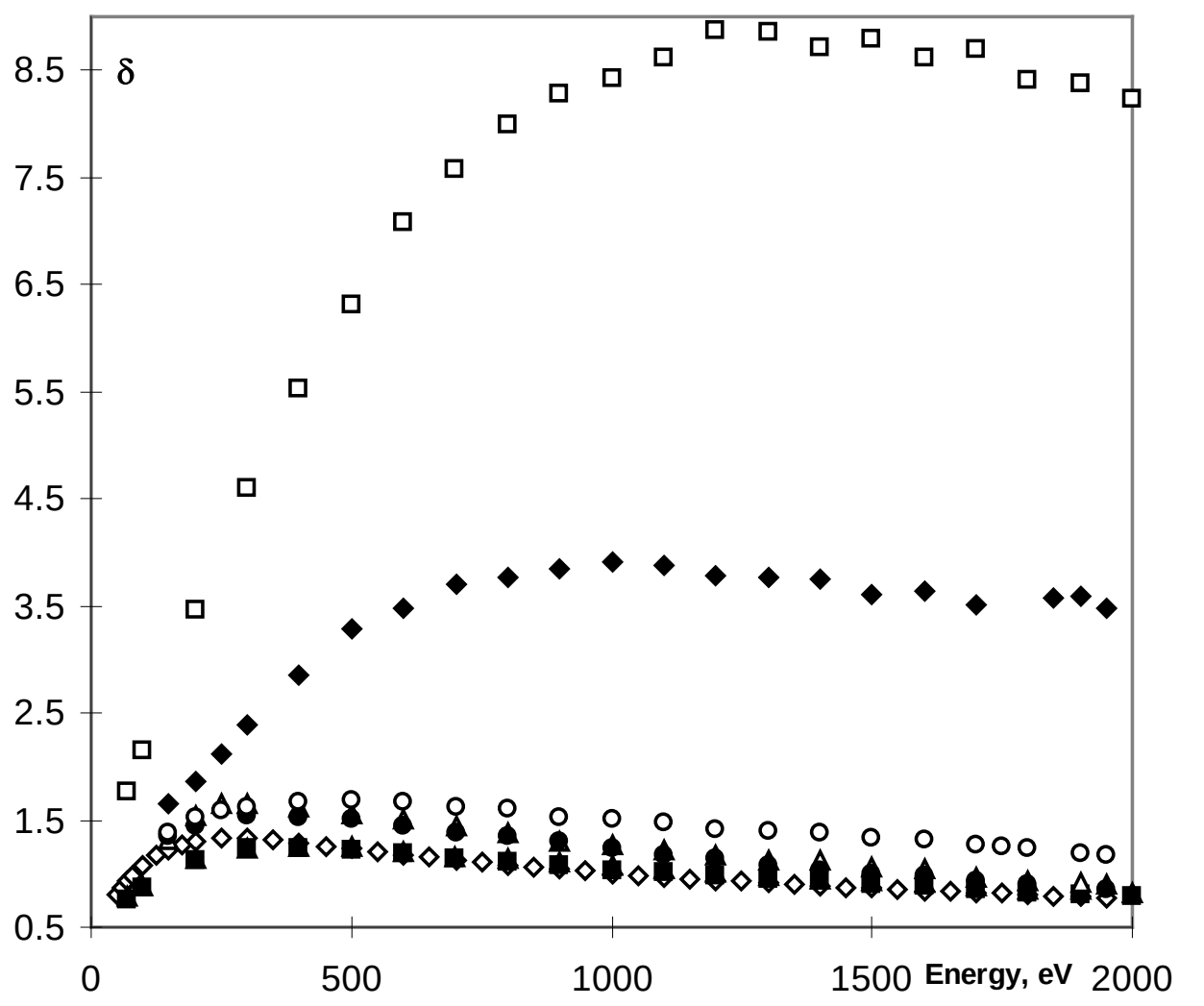

*Fig. 11. SEEC for titanium coated alumina samples after baking out at 350$^0$C*

The results of the SEEC measurements for these two series of titanium coated alumina samples in “as received” state and after baking out are presented correspondingly on Figures 10 and 11. Results obtained for the bulk of alumina used as a substrate for the samples are given as a reference. Also results for a titanium sample are shown for comparison.

*Chrome oxide and glass coated alumina samples.*

Sometimes it is interesting to have a coating on an insulator having low value of the SEEC and the same time high insulating properties. The glass seems to be a possible candidate corresponding these criteria. Two glass coated alumina samples have been prepared at DESY and chemical composition of the coating was $K_2O+SiO_2+Al_2O_3$. The samples were of thickness of 0.7mm.

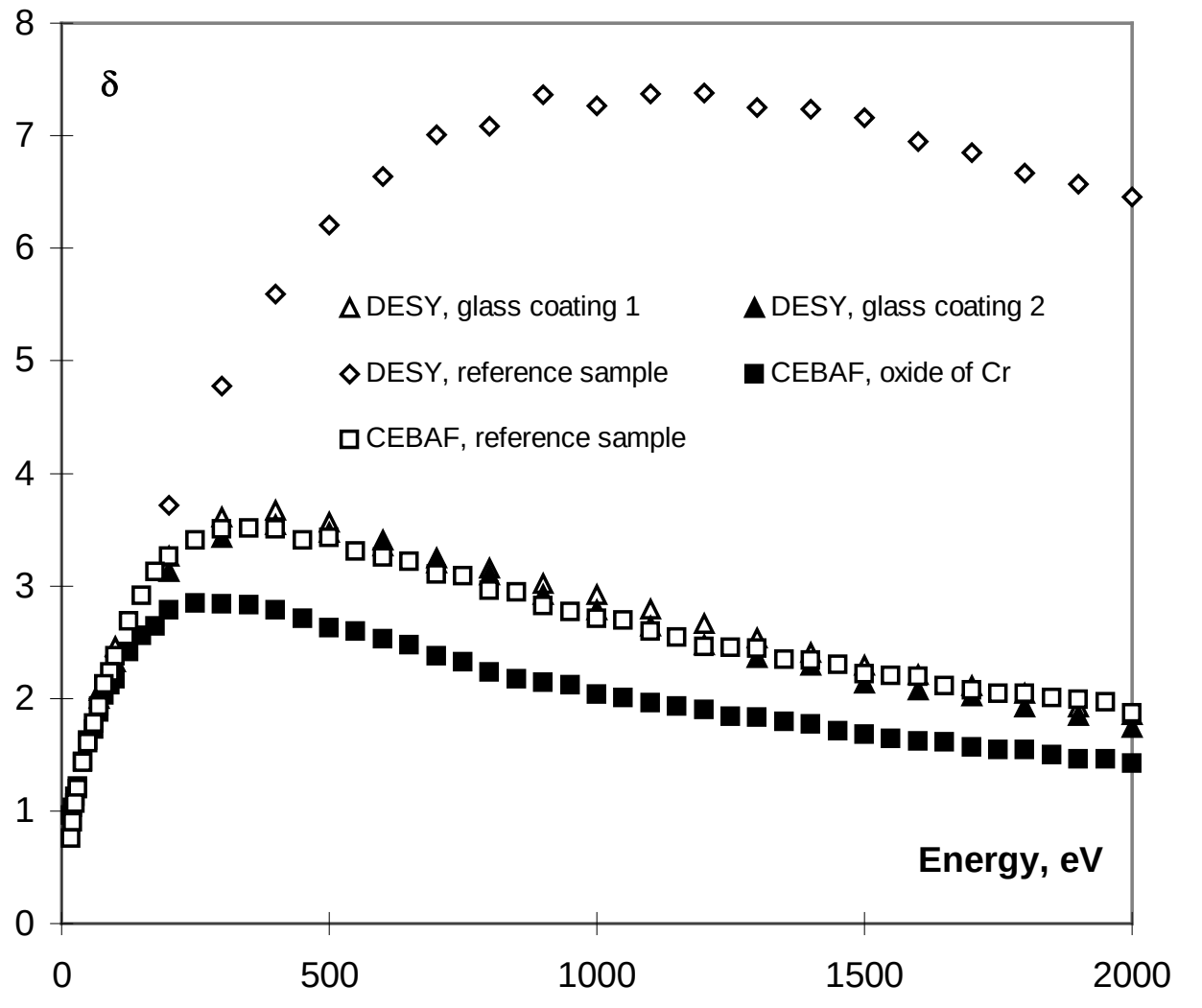

*Fig. 12. SEEC for chrome oxide and glass coated alumina samples in “as received” state.*

An alumina sample with chrome oxide coating has been prepared at CEBAF. The sample was of thickness of 0.2mm.

The Figures 12 and 13 present results of the SEEC measurements for glass and chrome oxide coated alumina samples in "as received" state and after baking out. Also, results obtained for the bulk of alumina used as a substrate for the samples are given as a reference.

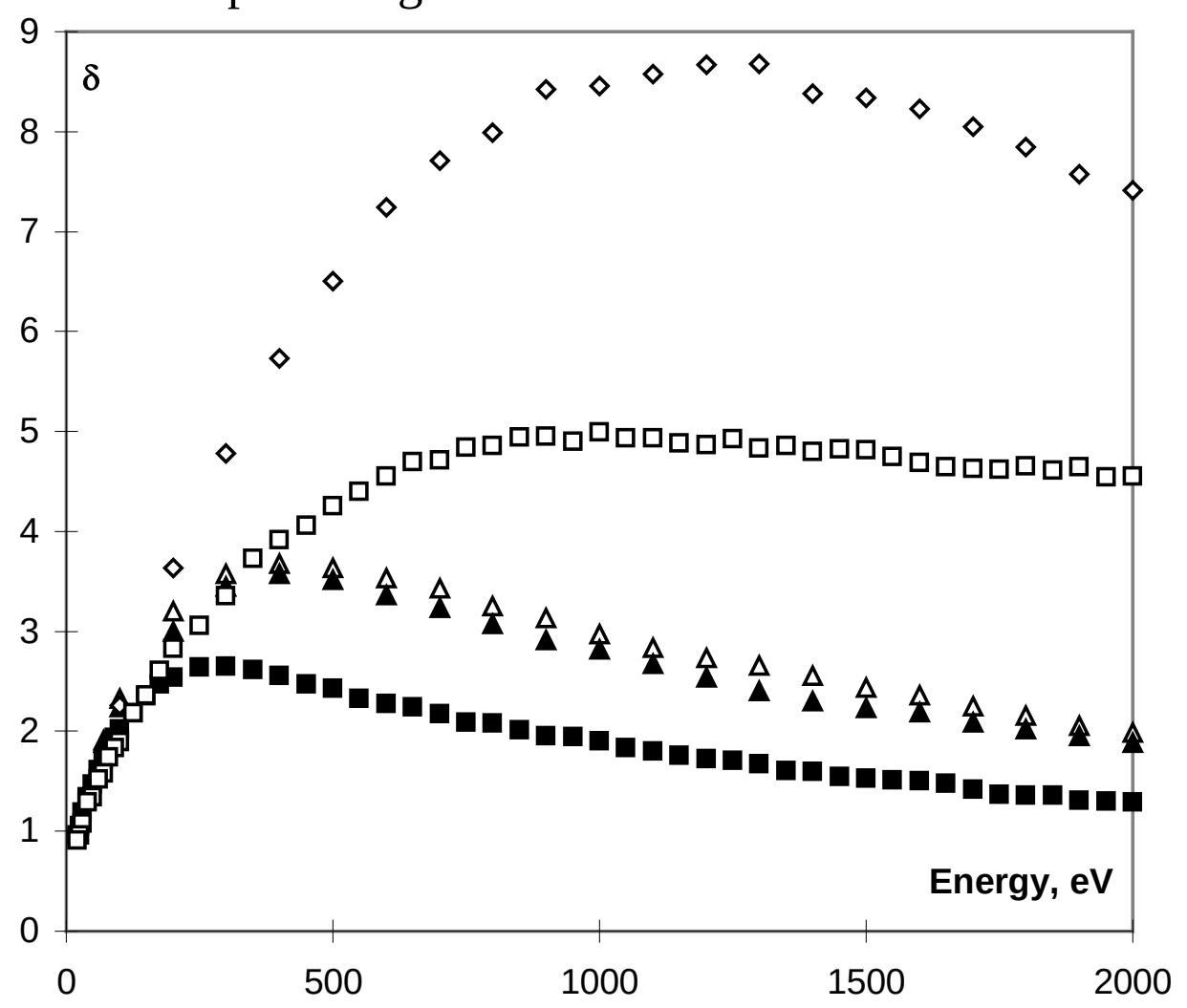

*Fig. 13. SEEC for chrome oxide and glass coated alumina samples after baking out at 350$^0$C*

## Discussion

The surface of any sample which has been subjected even a short air exposure before putting in a vacuum chamber, after pumping down until UHV conditions still remains contaminated with few monolayers of adsorbates. Such layers consist mainly of $H_2O$, CO and especially hydrocarbons. The baking at high temperatures yields to a surface close to the one of a pure material, i.e. almost free from adsorbed gases and some tightly bound species but still remaining an oxidized surface layer. Influence of the adsorbates on the SEEC can be observed here comparing its behavior in "as received" state and after baking out at 350$^0$C. Sometimes the change in SEEC is striking as for example for a pure alumina, the SEEC increases more than 2 times after baking out. Some samples as zyranox and glass on alumina do not exhibit any change and some samples as quartz and titanium coatings show more low values of the SEEC after baking out.

The increase of the SEEC after baking out can not be explained in the sense of production of secondary electrons and can be illustrated with the following example. It was found[11] that water frozen on a silver target at 77K exhibits a maximal value of the SEEC=2.3 at a coverage of about 200 monolayers and incident energy of 300eV. Taking into account an exponential character of energy loss in solid one can calculate that after having passed 10 monolayers of water a primary electron will loose only about 5% of its energy. Thus the production of the secondaries will be almost entirely defined by a substrate and will not be influenced by the adsorbate.

The presence a few monolayers of adsorbates can influence on the SEEC in the sense of production of the secondaries only in the case of a very low energy incident beam when incident electrons will loose a considerable part of energy during their travel via adsorbate or when the SEEC of an adsorbate is much higher than the one of a substrate. Since the similar as for water values have been found for $CO_2$[11] one has not to expect something very spectacular and too far from these values for other species making up an adsorbate. An exception could be allowed for rare gases condensed on a cold surface since the SEEC for them is expected to be very high and presence of few monolayers of a rare gas can drastically influence the SEEC of a substrate.

Therefore there should be another reason explaining influence of an adsorbate on SEEC and concerned with the escape mechanism of the secondary electrons. In order to escape into vacuum the secondary electrons must clear work function. It was already proven experimentally influence of work function on SEEC but the change in the SEEC was found not very big[12]. These experiments have been carried with metals having generally energy of the secondary electrons around 5eV. As it was measured for KCl[3] the maximum energy for secondary electrons was found to be 1eV. Such low values are much more influenced by the change in work function and this could explain the behavior of some investigated insulators. The removing the adsorbate from the surface of an insulator decreases work function and causes thus the escape of additional secondaries. In fact the lower energy of the secondary electrons the more the SEEC is influenced by the change in work function.

Impurities in the bulk of an insulator can also considerably change the SEEC. Results obtained for alumina samples of different purity vary from 5.7 to 10. A plot "Maximum SEEC -purity" exhibits a straight line what could be explained by the same kind of impurities for all these samples.

Sapphire and alumina have the same chemical composition but different lattice structure. The SEEC measurements for these materials do not show influence of lattice structure on SEEC.

Results obtained on titanium coatings are very close to the ones for the bulk titanium. It means that the thickness of the coatings is higher than the escape depth of the secondary electrons for the titanium. Such coatings are being successfully used at CERN and DESY as an effective antimultipactor coating for RF windows. The tested chrome oxide coating on alumina has exhibited results considerably higher of those obtained in SLAC[9,10]. Difference could be attributed either to another techniques of chrome oxide films preparation or to much less thickness of the coating which eventually could be less than the escape depth of the secondary electrons.

In general insulators are not simple chemical elements and most of them are metal compounds. This puts serious problems in creation of a solid theoretical basis explaining such phenomenon occurring in insulators as high electron yield, Malter effect etc. In this light, studies of the secondary electron emission from rare gases which are in fact, insulators could be an attractive research helping to find an answer to these questions.

## Acknowledgments

The authors would like to thank Mr. D. Bakker for his help in preparation of the beam driving electronics.